\documentclass[aps,preprint]{revtex4}
\usepackage{graphicx}
\usepackage{epsfig}
\usepackage[utf8]{inputenc}
\usepackage{float}
\DeclareUnicodeCharacter{200B}{}
\usepackage{xcolor}
\definecolor{DarkRed}{RGB}{139,0,0}
\definecolor{DarkBlue}{RGB}{0,0,139}
\definecolor{DarkGreen}{RGB}{0,100,0}

\begin{document}

\title{Horizon structure and extremal configurations of Kerr–Newman–anti--de Sitter black holes in  $f(R)$  gravity}
\author{\large Alikram N. Aliev}
\address{Department of Basic Sciences, Faculty of Engineering and Natural Sciences,
Maltepe University, 34857 Maltepe, Istanbul, T\"{u}rkiye}

\date{\today}

\begin{abstract}

Exploiting the correspondence between Kerr--Newman--(A)dS black hole solutions in general relativity and their constant-curvature counterparts in $f(R)$ gravity, we employ a unified framework to investigate the horizon structure and extremal configurations of these black holes, focusing primarily on the anti--de Sitter case. By solving the extremality conditions with respect to the squared rotation parameter $a^2$ and the inverse curvature scale $l^{-2}$, we obtain closed analytic relations that provide a transparent parametrization of the extremal branch in the $(a^2,\, l^{-2})$ parameter space. We determine the physically admissible domain of the rotating extremal branch and derive the corresponding critical charge, $q_{\max}=M/2$, at which the metric becomes singular and the branch reaches its critical rotation limit. For vanishing electric charge, the rotation parameter decreases monotonically along the extremal branch and remains bounded from below, implying the existence of a minimal rotation, $a_{\min}=3\sqrt{3}M/8$.
 The rotating extremal branch also exhibits a distinguished internal point at which the normalized rotational and curvature scales coincide. The inclusion of electric charge introduces an additional competing scale, broadening the rotating extremal branch while lowering the corresponding minimal rotation to $a_{\min}=M/4$ at the critical charge $q_{\max}=M/2$. Comparing these results with the corresponding de Sitter case, we show that extremality in AdS is constrained by a geometric endpoint, whereas the dS branch exhibits a considerably richer structure characterized by an ultra-extremal maximum and a subsequent decay toward the non-rotating limit.  Finally, we demonstrate that the algebraic mass constraint associated with the quartic horizon equation completely removes the real-root structure, yielding a class of solutions without physical horizons.

\end{abstract}

\maketitle

\newpage

\section{Introduction}

Despite its remarkable consistency and predictive power, general relativity  is not expected to represent the final classical theory of gravitation. It is widely believed that general relativity breaks down in regimes of high curvature, such as in the interior of black holes and in the early stages of cosmological evolution. This has led to sustained interest in extensions of general relativity, motivated by both theoretical developments, such as those arising in supergravity and string theory and observational challenges, including the need to account for dark energy and the late-time acceleration of the Universe \cite{stelle, birrel, vstar, perl}. Among the various modifications of general relativity, theories involving higher-order curvature invariants, most notably $f(R)$ gravity, are of particular interest. This theory represents one of the simplest and most consistent extensions of general relativity, in which the Einstein--Hilbert action is replaced by a nonlinear function of the Ricci scalar $R$. Remarkably, it leads to a rich phenomenology while retaining a relatively tractable mathematical structure.  In particular, $f(R)$ gravity admits a well-posed initial value formulation, making it suitable for numerical simulations of fully nonlinear dynamics. Taken together, these features have made $f(R)$ gravity as one of  the central subjects of modern investigations in modified gravity and cosmology over the past decades (see Refs.  \cite{star, soti, nojiri, cardoso}).

Black hole solutions in $f(R)$ gravity provide an important arena for exploring the physical properties and implications of this theory in the strong-field regime. Recent advances in observational black hole physics, such as the direct detection of gravitational waves from merging black holes in binary systems and the obtaining of the first horizon-scale images of the black holes at the centers of the M87 and Milky Way galaxies \cite{abbott1, akiyama1}, have significantly intensified efforts to test possible deviations from the predictions of general relativity. In particular, within $f(R)$ gravity and other alternative theories, considerable attention has been devoted to investigating the dynamical manifestation of additional scalar degrees of freedom. Such effects may appear, for instance, through scalarization phenomena in the inspiral--merger--ringdown waveforms of black hole binaries, which have no counterpart in general relativity \cite{lehner1, lehner2, pompili}.

Meanwhile, the construction of exact black hole solutions in $f(R)$ gravity is, in general,  a technically challenging problem, as the corresponding field equations derived from the nonlinear action involve up to fourth-order derivatives of the metric \cite{nojiri}. Nevertheless, a variety of static and spherically symmetric solutions, including Schwarzschild and Reissner--Nordström black holes, have been constructed in the literature \cite{maroto, son, odin1, tang}. Remarkably, imposing the constant-curvature condition $R = R_0$ plays a distinguished role, as it reduces the field equations to the Einstein form with an effective cosmological constant. This simplification allows one to construct a class of constant-curvature metrics for stationary and axisymmetric spacetimes, which, after an appropriate redefinition of the metric parameters, can be mapped onto the familiar Kerr--(A)dS and Kerr--Newman--(A)dS solutions of general relativity. This establishes a direct link between $f(R)$ gravity and the corresponding solutions in Einstein gravity \cite{larran, cembra}.

It is worth emphasizing that the black hole solutions considered here differ from those obtained in phenomenological dark-energy models \cite{kisel, bobo} (see also \cite{nojiri}). In the latter, dark energy is typically introduced as an effective matter source, leading to modified black hole geometries. In contrast, in the present framework the effective cosmological constant emerges entirely from the gravitational sector through the constant-curvature condition, without introducing additional matter degrees of freedom beyond the electromagnetic field. Therefore, although certain qualitative features of the horizon structure may appear similar, the solutions studied here correspond to exact constant-curvature electrovacuum configurations of $f(R)$ gravity rather than black holes supported by prescribed dark-energy sources.

In what follows, we restrict attention to constant-curvature black hole solutions of $f(R)$ gravity, characterized by $R = R_0$. The sign of $R_0$ distinguishes the geometries considered: negative values correspond to the anti--de Sitter case, whereas positive values describe the de Sitter case (equivalently, negative and positive cosmological constants, respectively).

In light of these developments, it is important to examine the horizon structure and extremal configurations of Kerr--Newman--(A)dS black holes in $f(R)$ gravity with constant curvature. Extremality conditions are typically associated with the degeneration of horizon roots and encode delicate balances among the physical parameters of the solution. In the presence of nonzero background curvature, these conditions acquire additional structure that competes with rotation and charge, leading to qualitatively new features compared to the asymptotically flat case. While extremal configurations of such black holes have been extensively studied in general relativity, a systematic analytic study of the extremal branches and their global behavior in the $(a^2, \,l^{-2})$ parameter space remains largely unexplored.

In our previous work \cite{an1}, we analyzed the horizon structure and extremal configurations of Kerr--Newman--de Sitter black holes in $f(R)$ gravity within a unified analytic framework, revealing a rich internal structure of the extremal branch. In particular, this structure is characterized by the emergence of an ultra-extremal configuration and the presence of an effective minimum of the rotation parameter before it decreases toward the non-rotating limit. In the present work, continuing this line of investigation, we turn to the anti--de Sitter case and examine the corresponding Kerr--Newman solutions in $f(R)$ gravity with constant curvature. Our analysis is based on the structure of the quartic horizon equation and its roots, expressed through compact analytic relations among the physical parameters. This approach enables a systematic classification of extremal configurations and the constraints imposed by the underlying algebraic structure and reveals new features of extremal Kerr--Newman--AdS black holes. It also provides a transparent comparison with their de Sitter counterparts and clarifies the role of negative background curvature as a scale competing with rotation and charge. 

The paper is organized as follows.  In Sec.~II, we present the field equations of $f(R)$ gravity and show that, under the constant-curvature condition $R = R_0$, they reduce to the Einstein form with an effective cosmological constant. Within this framework, we also discuss the Kerr--Newman--(A)dS family of solutions and their relation to the corresponding configurations in general relativity. 
 In Sec.~III, we analyze the structure of the quartic horizon equation and obtain an analytic characterization of the rotating extremal branch by deriving closed relations among the physical parameters. In particular, we determine its physically admissible domain and identify characteristic features such as the existence of a minimal rotation and the influence of electric charge on the branch geometry.  In Sec.~IV, we compare the extremal structure in AdS and dS spacetimes, highlighting the qualitative differences between the corresponding extremal branches. Finally, in Sec.~V, we examine the implications of the mass constraint for the root structure and show that it leads to a class of solutions without physical horizons. We conclude with a summary of the main results in Sec.~VI.

\section{Field equations and the Kerr--Newman--(A)dS metrics}

The field equations of $f(R)$ gravity coupled to a Maxwell field are derived by varying the action \cite{star, nojiri}
\begin{eqnarray}
S &=&  \int d^4 x \sqrt{-g} \left[\frac{1}{16 \pi G}\,\left (R + f(R) - 2 \Lambda \right) 
- \frac{1}{16 \pi} \,  F_{\mu\nu} F^{\mu\nu} \right] ,
\label{action}
\end{eqnarray}
with respect to the metric $g_{\mu\nu}$ and the vector potential $A_{\mu}$, where $f(R)$ is an arbitrary function of the Ricci scalar, $\Lambda$ denotes the cosmological constant and $F_{\mu\nu} = 2 \partial_{[\mu} A_{\nu]}$. This yields
\begin{eqnarray}
\label{eq1}
 && R_{\mu\nu} \left( 1 + f'(R) \right) - \frac{1}{2}\,  g_{\mu\nu} \left( R+f(R) - 2 \Lambda \right)  + \left( g_{\mu\nu} \, 
 \nabla^2 - \nabla_\mu \nabla_{\nu} \right) f'(R)    =  8\pi G T_{\mu\nu}\, ,  \\ [3mm] 
&& \nabla_{\nu} F^{\mu \nu}  =  0 \,. 
 \label{eq2}
\end{eqnarray}
Here $f'(R)= df/dR$, the operator $\nabla_{\mu}$ denotes covariant differentiation and $T_{\mu \nu}$ is the electromagnetic energy--momentum tensor satisfying $T_{\mu}^{\mu}=0$. The trace of Eq.~(\ref{eq1}) is given by
\begin{eqnarray}
3\, \nabla^{2} f'(R) + R \left(1 + f'(R) \right) &=& 2 \left( R + f(R) - 2\Lambda \right).
\label{treq1}
\end{eqnarray}

Next, we focus on the constant-curvature case, $R = R_0$. Substituting this condition into the trace equation yields
\begin{eqnarray}
R_0 &=& \frac{2\left[f(R_0) - 2\Lambda \right]}{f'(R_0) - 1} \,.
\label{constantR}
\end{eqnarray}
Using this relation together with the constant-curvature condition in (\ref{eq1}), we find that the field equations take the form
\begin{eqnarray}
 R_{\mu \nu}& = &\frac{1}{2}\,  g_{\mu \nu}\, \frac{f(R_0) - 2\Lambda}{f'(R_0)  - 1} +  \frac{8\pi G  T_{\mu \nu}}{1+f'(R_0)}\,.
\label{einform}
\end{eqnarray}
Clearly, this equation  can be written in the standard Einstein form of general relativity by introducing an effective cosmological constant,
\begin{eqnarray}
\lambda &=& \frac{1}{2}\, \frac{f(R_0) - 2\Lambda}{f'(R_0) - 1}
= \frac{R_0}{4} \,,
\label{effcosmcon}
\end{eqnarray}
together with an appropriate rescaling by the factor $1 + f'(R_0)$ of the Newtonian constant (or, equivalently, of the energy--momentum tensor $T_{\mu\nu}$). For further details, see Ref. \cite{an1}.

This observation suggests the existence of a correspondence between a class of stationary and axisymmetric spacetimes in general relativity and their constant-curvature counterparts in $f(R)$ gravity. In this framework, the Kerr--Newman--(A)dS family of metrics in general relativity can be mapped into the corresponding solutions of $f(R)$ gravity after appropriate rescalings of the metric parameters \cite{larran, cembra}. Consequently, within a unified framework of both general relativity and $f(R)$ gravity, the Kerr--Newman--(A)dS metrics describing rotating charged black holes in Boyer--Lindquist coordinates can be written in the form
\begin{eqnarray}
ds^2 & = & -{{\Delta_r}\over {\Sigma}} \left(\,dt - \frac{a
\sin^2\theta}{\Xi}\,d\phi\,\right)^2 + {\Sigma \over \Delta_r}
dr^2 + {\Sigma \over \Delta_{\theta}}\,d\theta^{\,2} +
\frac{\Delta_{\theta}\sin^2\theta}{\Sigma} \left(a\, dt -
\frac{r^2+a^2}{\Xi} \,d\phi \right)^2,
\nonumber\\
\label{4knads}
\end{eqnarray}
where the metric functions
\begin{eqnarray}
\Delta_r &= &\left(r^2 + a^2\right)\left(1 + \frac{ r^2}{l^2}\right)
- 2 M r + \frac{Q^2}{1+f'(R_0)}\,, \qquad
\Sigma = r^2+ a^2 \cos^2\theta \,,
\nonumber \\[2mm]
\Delta_\theta & = & 1 - \frac{ a^2}{l^2}
\cos^2\theta\,, \qquad
\Xi = 1 - \frac{ a^2}{l^2} \,,
\label{metfunct}
\end{eqnarray}
and the associated vector potential is given by
\begin{equation}
A= -\frac{Q \,r}{\Sigma}\left(dt- \frac{a
\sin^2\theta}{\Xi}\,d\phi \right).
\label{potform}
\end{equation}
Here $M$, $a$ and $Q$ denote the mass, rotation  and charge parameters, respectively \cite{an2}. For convenience, we introduce the curvature scale $l$, determined by the negative AdS background curvature through
\begin{eqnarray}
l^{-2} = -\frac{R_0}{12} = -\frac{\Lambda}{3}\,,
\label{ll}
\end{eqnarray}
where the relation $R_0 = 4\Lambda$, as given in Eq.~(\ref{effcosmcon}), connects the constant scalar curvature $R_0$ to the cosmological constant $\Lambda$. It then follows that positive dS curvature is obtained through the formal replacement $ l^{2} \rightarrow -\, l^{2}.$ Thus, the introduction of the curvature scale $l$ allows one to treat both (A)dS and $f(R)$ black holes within a unified framework.

We note that the squared charge parameter appearing in the metric function $\Delta_r$ is multiplied by the constant factor $(1+f'(R_0))^{-1}$. This originates from the corresponding rescaling of the energy--momentum tensor $T_{\mu\nu}$ in Eq.~(\ref{einform}) (instead of the Newtonian constant $G$, which is set equal to one). For later convenience, we introduce the rescaled charge parameter $q$ through
\begin{eqnarray}
q^2 = \frac{Q^2}{1+f'(R_0)}\,,
\label{chargepara}
\end{eqnarray}
which reduces to the standard Kerr--Newman charge in the limit $f'(R_0)\rightarrow 0$.

It is also important to note that the quantity $\Xi$  in  (\ref{metfunct}) removes conical singularities along the axis of symmetry.  It then follows that the rotation parameter is constrained by the condition $a^2 \leq l^2$. In the critical rotation limit $a^2=l^2$, corresponding to $\Xi=0$, the metric becomes singular. In this case, the angular velocity of the horizon takes the simple form
\begin{equation}
\omega_{H}=\frac{1}{l}\,,
\label{crit}
\end{equation}
independently of the horizon radius \cite{haw, an2}. This coincides with the angular velocity of the Einstein universe at the AdS boundary, implying that the boundary rotates at the speed of light, $ v=\omega l \rightarrow 1, $ as $a^2 \rightarrow l^2$. 

\section{Horizon structure and extremal configurations}

In this section, we analyze the horizon structure and extremal configurations of the solution (\ref{4knads}) by examining the roots of the horizon equation, $\Delta_r = 0$, which can be written  in the  form
\begin{eqnarray}
r^4 + a_2\, r^2 + a_1\, r + a_0 &=& 0 ,
\label{quartic1}
\end{eqnarray}
where the coefficients are given by
\begin{eqnarray}
a_2 &=& l^2 + a^2 , \qquad
a_1 = - 2 M l^2 , \qquad
a_0 = l^2 \left(a^2 + q^2\right) .
\label{constcoef}
\end{eqnarray}

The root structure of this quartic equation was earlier analyzed in \cite{an2}. In particular, it was shown that the equation admits two real roots, where the larger root $r_{+}$ determines the location of the black hole event horizon, while the smaller root $r_{-}$ corresponds to the inner (Cauchy) horizon.  For completeness, we briefly recall the explicit expressions for these roots, obtained in \cite{an2} using Cardano's method \cite{astegun},
\begin{eqnarray}
r_{+} &=& \frac{1}{2}\left(X + Y \right), \qquad
r_{-} = \frac{1}{2}\left(X - Y \right),
\label{horizons}
\end{eqnarray}
where
\begin{eqnarray}
X &=& \sqrt{u - l^2 - a^2}, \qquad
Y = \sqrt{-u - l^2 - a^2 + \frac{4 M l^2}{X}}.
\end{eqnarray}
Here $u$ denotes a real solution of the associated resolvent cubic equation,
\begin{eqnarray}
u &=& \frac{l^2 + a^2}{3}
+ \frac{l^{4/3}\left(M_{1e}^2 - M_{2e}^2 \right)^{2/3}}
{\left(2 N^2 - M_{1e}^2 - M_{2e}^2 \right)^{1/3}}
+ l^{4/3} \left(2 N^2 - M_{1e}^2 - M_{2e}^2 \right)^{1/3},
\label{resolventr}
\end{eqnarray}
with
\begin{equation}
N^2 = M^2 + \sqrt{\left(M^2 - M_{1e}^2\right)\left(M^2 - M_{2e}^2\right)}.
\label{censor1}
\end{equation}
We have also introduced two extremal mass parameters, given by
\begin{eqnarray}
M_{1e}^2 &=& \frac{l^2}{54}\left(\zeta + \eta^3 \right), \qquad
M_{2e}^2 = \frac{l^2}{54}\left(\zeta - \eta^3 \right),
\label{exmasses}
\end{eqnarray}
where
\begin{equation}
\zeta = \left(1+\frac{a^2}{l^2}\right)\left[\frac{36(a^2+q^2)}{l^2}-\left(1+\frac{a^2}{l^2}\right)^2\right], \qquad
\eta = \left[\left(1+\frac{a^2}{l^2}\right)^2+ \frac{12(a^2+q^2)}{l^2}\right]^{1/2}.
\label{zetaeta}
\end{equation}

In the limit of vanishing cosmological constant, $l \to \infty$, the mass parameter $M_{1e}$ reduces to the extremal mass of Kerr--Newman black holes, $M_{1e}^2 \to a^2 + q^2$, while $M_{2e}^2 \to -\infty$ and therefore has no physical significance. For further insight, we expand the roots (\ref{horizons}) in powers of $1/l^2$. Retaining terms up to first order, we obtain
\begin{eqnarray}
r_{+} &=& \tilde{r}_{+} -\frac{\tilde{r}_{+}^{2}}{2\, l^2}
\,\frac{2 M \tilde{r}_{+} - q^2}{\tilde{r}_{+}-M },\\[2mm]
r_{-} &=& \tilde{r}_{-} -\frac{\tilde{r}_{-}^{2}}{2\, l^2}
\,\frac{2 M \tilde{r}_{-} - q^2}{\tilde{r}_{-}-M },
\label{limits1}
\end{eqnarray}
where
\[
\tilde{r}_{\pm} = M \pm \sqrt{M^2 - a^2 - q^2}.
\]
We observe that the event horizon satisfies $r_{-} < r_{+} < \tilde{r}_{+}$. Thus, in the presence of negative scalar curvature (or, equivalently, a negative cosmological constant), both the outer and inner horizons are shifted inward relative to their Kerr--Newman counterparts in asymptotically flat spacetime. It is worth emphasizing that the black hole configurations considered here are not regular, since Kerr--(Newman)--de Sitter spacetimes generically possess a curvature singularity at the center. For an interesting analysis of multi-horizon configurations in a class of regular black holes, see \cite{odin2}.

Building on the above results, we investigate the extremal configurations of Kerr--Newman black holes in $f(R)$ gravity with negative scalar curvature. These arise when the inner and outer horizons coincide, $r_{+}=r_{-}$, corresponding to a double root of the horizon equation.
This condition is determined by the simultaneous equations
\begin{eqnarray}
\Delta_r &=& 0\,, \qquad
\frac{d\Delta_r}{dr} = 0\,,
\label{simultextr}
\end{eqnarray}
which describe the merging of the two horizons into a single extremal configuration.

In what follows, we focus on the rotating extremal branch obtained from the above degeneracy conditions. The bounds and limiting values derived below are therefore understood as properties of this branch.

Solving this system, using Eq.~(\ref{quartic1}), we find that the extremal mass is given by $M = M_{1e}$, in exact agreement with that given in  Eq.~(\ref{exmasses}), while the corresponding extremal horizon radius is
\begin{eqnarray}
r_{-} & = &  r_{+} =\frac{l}{\sqrt{6}} \left(\eta - 1 - \frac{a^2}{l^2}\right)^{1/2}.
\label{exhorizon}
\end{eqnarray}

A particularly useful reformulation is obtained by solving the system of Eqs.~(\ref{simultextr}) instead with respect to the squared rotation parameter $a^2$ and the inverse square of the curvature scale $l^{-2}$. This yields closed analytical expressions, which provide a transparent parametrization of the extremal configurations in terms of a one-dimensional curve in the parameter space $(a^2,\, l^{-2})$. This representation makes the structure of the extremal solutions manifest. It is straightforward to show that the solutions to this system can be written as
\begin{eqnarray}
\label{aaq}
a^2 &=& \frac{M r - 2 r^2 - q^2 + \sqrt{Z}}{2}\,, \\[2mm]
l^{-2} &=& \frac{q^2 - M r - 2 r^2 + \sqrt{Z}}{2 r^4}\,,
\label{llq}
\end{eqnarray}
where
\begin{equation}
Z = M r^2 (M + 8 r) - q^2 \left(4 r^2 + 2 M r - q^2\right).
\label{Z}
\end{equation}
These expressions provide a closed analytic parametrization of the extremal configurations in terms of the horizon radius $r$. We further note that, under the replacement $l^{-2} \rightarrow -\,l^{-2}$, they are in exact agreement with the corresponding results of \cite{an1}.

As discussed in Sec.~II, the rotation parameter must satisfy $ a^2 \leq l^2 $, with equality corresponding to the critical limit $ a^2 = l^2 \, (\Xi = 0)$, at which the spacetime metric becomes singular. This limit corresponds to the maximal rotation allowed by the AdS background. This condition, in turn, determines the limiting size of the extremal black hole.  \, Using Eqs.~(\ref{aaq}) and (\ref{llq}), we impose the relation $ a^2 l^{-2}-1=0 $ and solve for the critical horizon radius $\tilde r_{e}$. This yields
\begin{eqnarray}
\tilde r_{e} &=& \frac{3 M^2 + 4 q^2 + \sqrt{(M^2 - 4 q^2)(9 M^2 - 4 q^2)}}{16 M}\,,
\label{critical1}
\end{eqnarray}
from which it follows that the electric charge is bounded by
\begin{eqnarray}
q_{\rm max} &=& \frac{M}{2}\, .
\label{limitingq}
\end{eqnarray}
This result defines the boundary of the physically admissible rotating extremal branch considered here. It should not be interpreted as a global upper bound on the charge of all extremal Kerr--Newman--AdS configurations; rather, it is the limiting value at which this branch reaches the critical rotation surface $a^2=l^2$ and the metric becomes singular. We note that the reality of the square root in Eq.~(\ref{critical1}) formally allows a second bound, $q = 3M/2$. However, this branch leads to negative values of $a^2$ through Eq.~(\ref{aaq}) and is therefore excluded from the physically admissible parameter space.

Next,  we analyze separately the uncharged $(q=0)$ and charged $(q \neq 0)$ cases for clarity. 

\subsection{The uncharged case, $q=0$}

In this case, Eqs.~(\ref{aaq}) and (\ref{llq}) simplify to
\begin{eqnarray}
\label{aa0}
a^2 &=& \frac{1}{2}\, r \left( M - 2 r + \sqrt{M}\sqrt{M + 8 r} \right),  
\\[2mm]
l^{-2} &=& \frac{- M - 2 r + \sqrt{M}\sqrt{M + 8 r}}{2 r^3}\,.
\label{ll0}
\end{eqnarray}
We see that the extremal value  of the squared rotation parameter $a^2$  and the inverse square of the curvature radius $ l^{-2} $ explicitly depends on the horizon location. This behavior arises from the presence of a nonzero negative scalar curvature $ R_0$ (or, equivalently, a negative cosmological constant $\Lambda$). Consequently, only in the asymptotically flat limit $l\to\infty$ one recovers the familiar Kerr extremality bound  $ a_{\rm max}=M $ with the corresponding extremal horizon located at  $ r=M $.  On the other hand, in the critical limit $a^2 = l^2$ , using Eq.~(\ref{critical1}) with $q=0$, we find that the limiting radius of the extremal black hole is
\begin{eqnarray}
\tilde r_{e} &=& \frac{3}{8}\,M \, .
\label{critical}
\end{eqnarray}
The corresponding value of the rotation parameter is then 
\begin{eqnarray}
a^2 &=& 3\, \tilde r_{e}^2, \qquad 
a = \sqrt{3}\,\tilde r_{e} \simeq 0.649\,M .
\label{aa01}
\end{eqnarray}
We conclude that  the extremal branch is defined for $
r \in \left(\frac{3M}{8},\, M\right)$.  At \(r=M\), the function \(a^2(r)\)  attains its asymptotically flat Kerr extremal limit $ a^2 = M^2 $. As  \(r\) decreases, \(a^2(r)\) decreases monotonically, reaching its minimum value given in Eq.~(\ref{aa01})
at the radius specified in Eq.~(\ref{critical}), where the extremal branch ends. Thus, along this rotating extremal branch, the rotation parameter is bounded from below, implying the existence of a branch-dependent minimal rotation. This statement does not refer to the full extremal solution space, which also contains the non-rotating charged limit.  In contrast, the function \(l^{-2}(r)\) increases monotonically from zero at \(r = M\) toward the endpoint of the extremal branch, defined by the radius given in Eq.~(\ref{critical}).  Along this curve, it is useful to introduce the dimensionless quantities
\begin{eqnarray}
\mathcal{A}=\frac{a^2}{M^2},\qquad 
\mathcal{L}=\frac{M^2}{l^2}.
\label{normalized}
\end{eqnarray}
The condition $\mathcal{A}=\mathcal{L}$ then selects a distinguished point at which the  normalized  rotational and curvature scales coincide. A numerical solution gives
\begin{eqnarray}
r &\simeq& 0.561\,M, \qquad
\frac{a^2}{M^2}= \frac{M^2}{l^2} \simeq 0.623
\qquad (a \simeq 0.789\,M).
\label{physroot2}
\end{eqnarray}
\begin{figure}[h!]
\centering
\begin{tabular}{cccc}
\epsfig{file=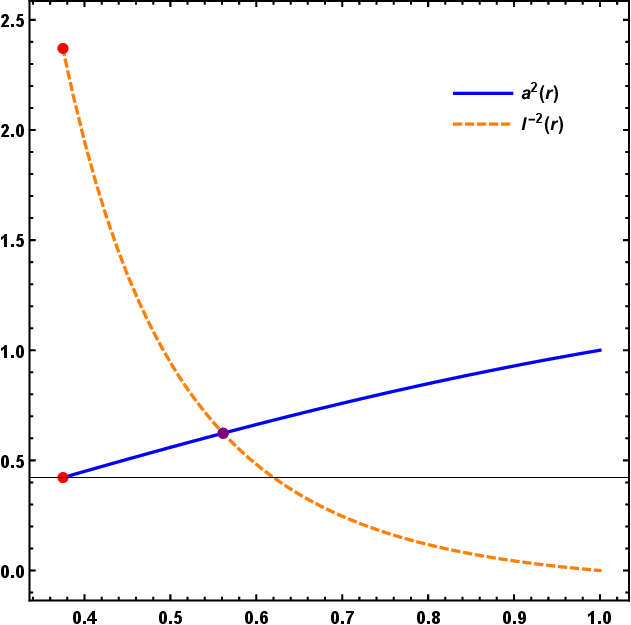,width=0.44\linewidth,clip=} &&&
\epsfig{file=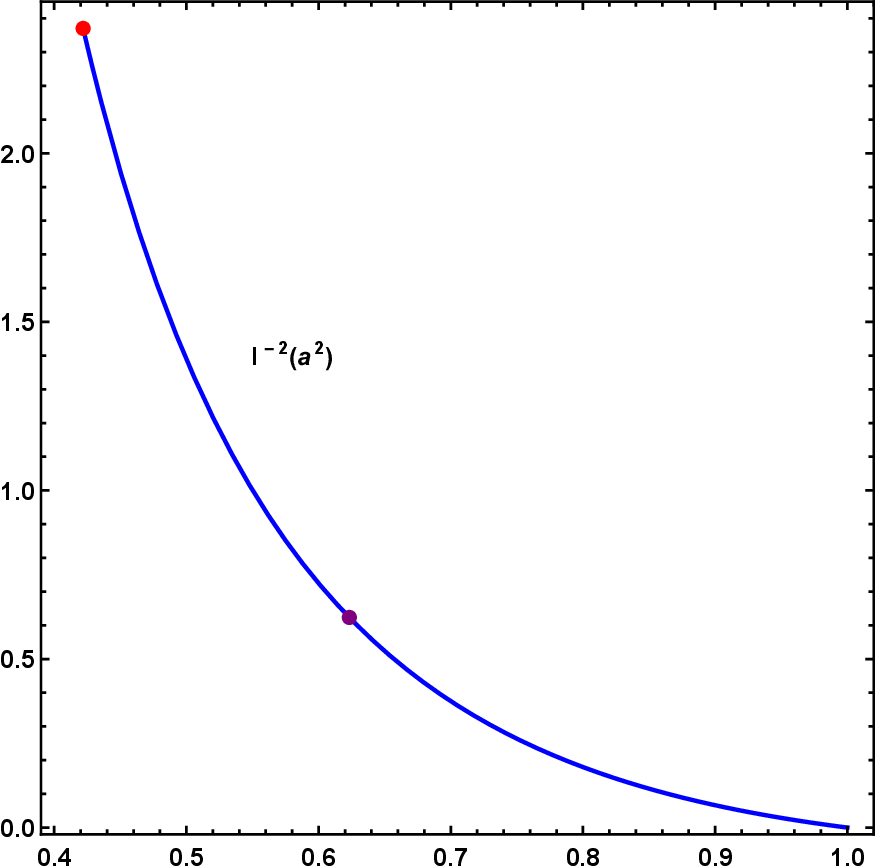,width=0.44\linewidth,clip=}
\end{tabular}
\caption{The left panel shows the dependence of the squared extremal rotation parameter $a^2$ and the inverse square of the curvature scale $l^{-2}$ on the horizon radius $r$. The function $a^2(r)$ decreases monotonically and reaches its minimum at the geometric endpoint $r = 3M/8$ (red point), while $l^{-2}(r)$ increases monotonically from zero at $r = M$ toward the same endpoint. The two curves intersect at a distinguished point (purple point), where the normalized rotational and curvature \mbox{scales} coincide. This point marks an intrinsic balance between the effects of rotation and background curvature. The right panel shows $l^{-2}$ as a function of $a^2$, illustrating that this intersection reflects an intrinsic feature of the extremal branch rather than a boundary or singular configuration.  In the figures, all quantities are measured in units of the mass parameter $M$.}%
\end{figure}%

\vspace{-0.2 cm}

It is important to note that this condition does not follow from the extremality equations themselves and therefore does not define a boundary or singular configuration. In contrast, the geometric constraint $ \Xi = 1 - a^2/l^2 = 0 $ corresponds to a true limiting point of the extremal branch, where the metric becomes singular and the branch  reaches its endpoint.  Thus, the intersection point identified above should instead be interpreted as a distinguished point associated with the matching of rotational and curvature scales.  In this sense, the extremal curve exhibits two qualitatively distinct types of special points: a boundary point associated with the breakdown of the geometry  and an internal point associated with the matching of intrinsic scales.  The results of this analysis are displayed in Fig.~1.

\vspace{1 cm}

\subsection{The charged case, $q \neq 0$}

We now turn to the charged case and continue the analysis of the rotating extremal branch introduced above. The presence of a nonvanishing electric charge modifies the structure of this branch.  In this case, the functions $a^2(r)$ and $l^{-2}(r)$ are still determined by Eqs.~(\ref{aaq}) and (\ref{llq}), but exhibit a more intricate dependence on the horizon radius due to the additional contribution of the charge parameter $q$. As a result, the interplay between rotation, curvature  and charge leads to a deformation of the extremal branch in parameter space.

In the limit of vanishing background curvature, $l \to \infty$, Eq.~(\ref{llq}) implies that the horizon radius reduces to $r = M$. Substituting this into Eq.~(\ref{aaq}), one recovers the familiar extremality condition $M^2 = a^2 + q^2$ for the Kerr--Newman black hole. For nonvanishing charge, the critical condition $a^2 = l^2$ determines the limiting size of the extremal horizon given in Eq.~(\ref{critical1}). 
In particular,  for the critical charge of the rotating extremal branch in Eq.~(\ref{limitingq}), this yields
\begin{eqnarray}
\tilde r_{e} &=& a = \frac{M}{4}\,.
\label{aarr}
\end{eqnarray}
Comparing this result with the uncharged case, where the corresponding endpoint is located at $ \tilde r_{e} = 3M/8$, we observe that the presence of electric charge reduces the size of the extremal horizon.  At the same time, the minimum value of the rotation parameter along this rotating extremal branch is also reduced relative to the uncharged case in the critical limit $a^2=l^2$.  Moreover, the rotating extremal branch is enlarged, extending from $r = M/4$ up to $r = M$. 
Thus, for the critical value of the electric charge, the extremal branch is defined over the interval $r \in \left(\frac{M}{4},\, M\right)$, indicating that the inclusion of charge both lowers the maximal rotation and broadens  the range of the rotating extremal branch.

We now focus  on the critical value of the  charge, $q_{max} = M/2$. In this case, at $r = M$, the function $a^2(r)$ attains its Kerr--Newman extremal value $a^2 = 3M^2/4$, and then decreases monotonically to its minimum value $a^2 = M^2/16$ at the endpoint of the extremal branch specified in Eq.~(\ref{aarr}). At the same time, the function $l^{-2}(r)$ increases monotonically from zero at $r = M$ toward the same endpoint, where the extremal branch terminates.  Along these curves, there exists an  intersection point determined by the condition  $\mathcal{A}=\mathcal{L}$, as defined in  Eq.~(\ref{normalized}).
 Using Eqs.~(\ref{aaq}) and (\ref{llq}), one finds numerically
\begin{eqnarray}
r &\simeq& 0.634\,M, \qquad
\frac{a^2}{M^2}= \frac{M^2}{l^2} \simeq 0.458
\qquad (a \simeq 0.677\,M).\nonumber
\label{physrootq}
\end{eqnarray}
\begin{figure}[h!]
\centering
\begin{tabular}{cccc}
\epsfig{file=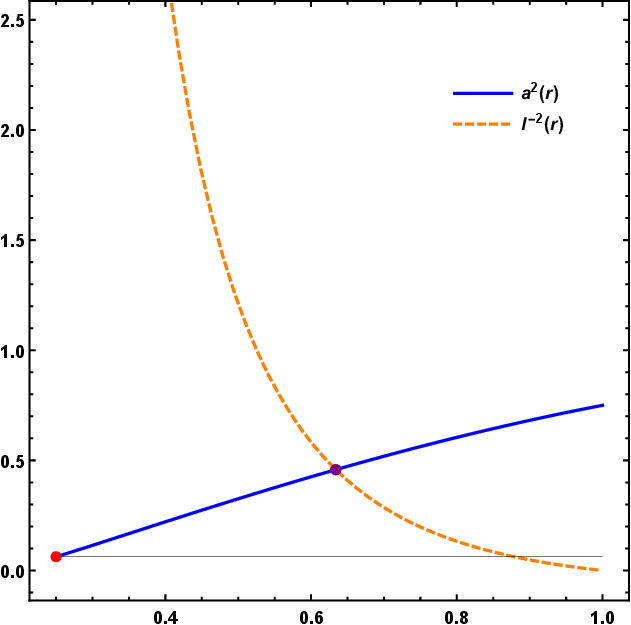,width=0.44\linewidth,clip=} &&&
\epsfig{file=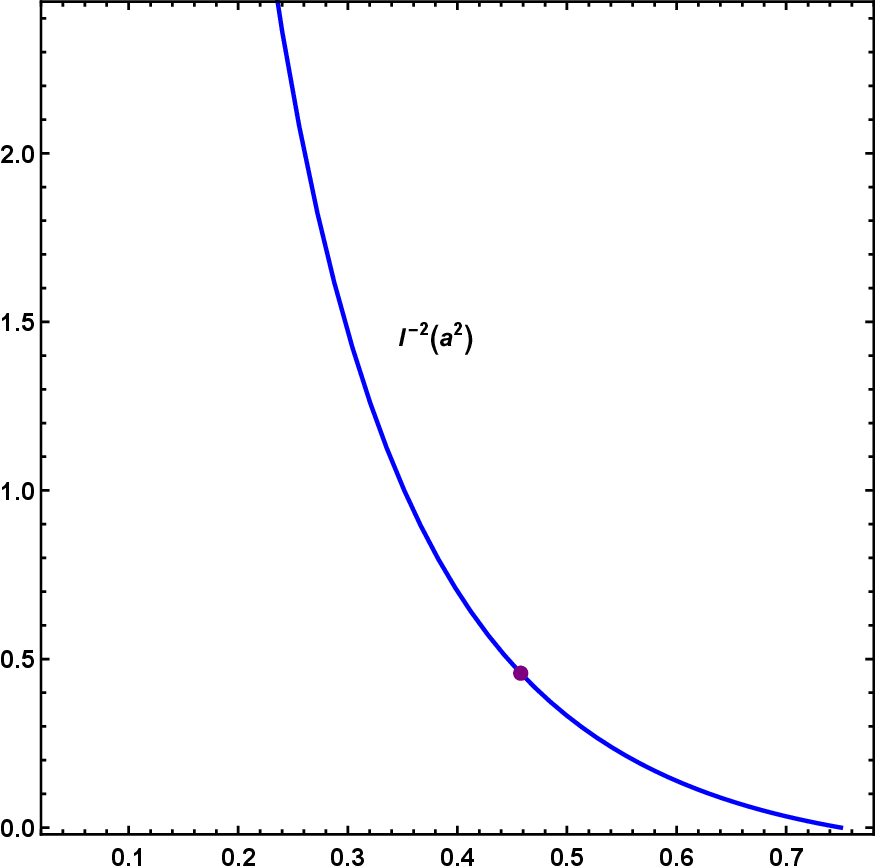,width=0.44\linewidth,clip=}
\end{tabular}
\caption{The left panel displays the behavior of the squared extremal rotation parameter $a^2$ and the inverse square of the curvature scale $l^{-2}$ as functions of the horizon radius $r$ for  the critical value of the charge $q_{max} = M/2$. In contrast to the uncharged case, the inclusion of electric charge extends  the rotating extremal branch from $r \in (3M/8,\,M)$ to $r \in (M/4,\,M)$, while shifting the intersection point to larger radii and reducing the corresponding rotation parameter. The function $a^2(r)$ decreases monotonically from its Kerr--Newman extremal value at $r = M$ to its minimum at the endpoint $r = M/4$, whereas $l^{-2}(r)$ increases from zero toward the same limiting point.  The two curves intersect at a distinguished point (purple point), whose location is shifted toward larger horizon radii by the presence of electric charge. The right panel shows $l^{-2}$ as a function of $a^2$, demonstrating that this point does not correspond to a limiting configuration, but instead  marks an intrinsic feature of the rotating extremal branch, where the effects of curvature and rotation become comparable."
The endpoint (red point) lies outside the displayed vertical range in this panel. All quantities in the figures are expressed in units of the mass parameter $M$. }%
\end{figure}%
Comparing this result with the uncharged case, where the corresponding intersection occurs at $r \simeq 0.561\,M$, we observe that the inclusion of electric charge shifts the intersection point to larger values of the horizon radius, while the corresponding rotation parameter is reduced. This indicates that the presence of charge enlarges the characteristic scale of the extremal configuration at the expense of rotation. This behavior reflects a balance between rotation, curvature and charge, with the extremal configuration adjusting to maintain equilibrium among these competing effects. Details of this analysis are illustrated in Fig.~2 .

To summarize, within the rotating extremal branch considered here, electric charge broadens the physically admissible domain while simultaneously lowering the corresponding rotation parameter. As a consequence, the geometry of the extremal branch is qualitatively modified, revealing the combined influence of rotation, curvature and electric charge. We also note that the results presented here are in qualitative agreement with those obtained in \cite{cembra, jiri}, which were derived primarily through numerical analyses of the corresponding highly cumbersome equations.

\section{Comparison of Extremal Structure in AdS and dS Cases}

We consider constant-curvature solutions of $f(R)$ gravity with $R = R_0$, where the AdS and dS cases correspond to negative and positive values of $R_0$ (or, equivalently, to negative and positive cosmological constants), respectively. For the AdS branch, the curvature scale is defined by Eq.~(\ref{ll}), so that $l^{-2} > 0$. The corresponding dS expressions are obtained by reversing the sign of the curvature contribution, equivalently by the formal replacement $l^{-2} \to -\,l^{-2}$ in the AdS formulas.  A detailed analysis of the horizon structure and extremal configurations for Kerr--Newman--de Sitter black holes was presented in \cite{an1}. Here, we compare those results with the corresponding Kerr--Newman--AdS case discussed above.  Although both cases are described within the same analytic framework, their extremal structures differ significantly because of the opposite signs of the background curvature.

{\it (i)\ The anti--de Sitter case.} As discussed above, for vanishing electric charge, $ q=0$, the extremal branch is defined for $r \in (3M/8,\,M)$. At $r = M$, one recovers the familiar extremal Kerr limit $a^2 = M^2$. As $r$ decreases, the function $a^2(r)$ decreases monotonically, reaching its minimum value $a^2 = 27 M^2/64$ at the endpoint $r = 3M/8$. This point coincides with the geometric bound $(\Xi = 0)$, at which the metric becomes singular and the extremal branch ends. Thus, in the AdS case, the extremal rotation parameter is bounded from below by a geometric endpoint.  Moreover, along the extremal branch there exists a distinguished internal point at which the normalized rotational and curvature scales coincide. This point does not correspond to a singularity; rather, it identifies a scale-matching configuration where the rotational and curvature contributions become of  the same order of magnitude.

{\it (ii)\ The de Sitter case.} As shown in \cite{an1}, the extremal branch exhibits a qualitatively different structure. The function $a^2(r)$ develops a maximum at
\begin{eqnarray}
\label{rultra}
r & =& \frac{3+2\sqrt{3}}{4}\,M, \,\,\,\,  with  \,\,\,\,  a^2 = \frac{3}{16}\left(3+2\sqrt{3}\right) M^2,
\end{eqnarray}
which may be interpreted as an ultra-extremal configuration. As $r$ increases further, $a^2(r)$ decreases and vanishes at $r = 3M$, corresponding to the Schwarzschild--de Sitter limit. Along this right branch, there exists a distinguished point where the curves $a^2(r)$ and $l^{-2}(r)$ intersect. This point does not mark an endpoint, as the branch extends up to $r = 3M$. Rather, it should be interpreted as an internal scale-matching point at which the rotational and curvature scales become comparable along the decaying branch. In this sense, it also represents an effective minimum of the rotation parameter, beyond which the solution rapidly approaches the non-rotating limit $a \to 0$.

Taken together, these results show that the extremal manifolds in AdS and dS spacetimes exhibit qualitatively different internal structures. In the AdS case, the rotation parameter is bounded from below by a geometric endpoint at which the rotating extremal branch reaches its limiting point. In contrast, the dS extremal branch admits a maximal (ultra-extremal) rotation and, along its right branch, an effective minimum before approaching the non-rotating limit. While the AdS extremal branch terminates at a boundary, the dS branch interpolates between a maximum and vanishing rotation, thereby exhibiting a richer internal structure. These features are illustrated in Fig.~3, where the left panel shows $a^2(r)$, while the right panel displays both $a^2(r)$ and $l^{-2}(r)$, including the corresponding intersection points.
\begin{figure}[h!]
\centering
\begin{tabular}{cccc}
\epsfig{file=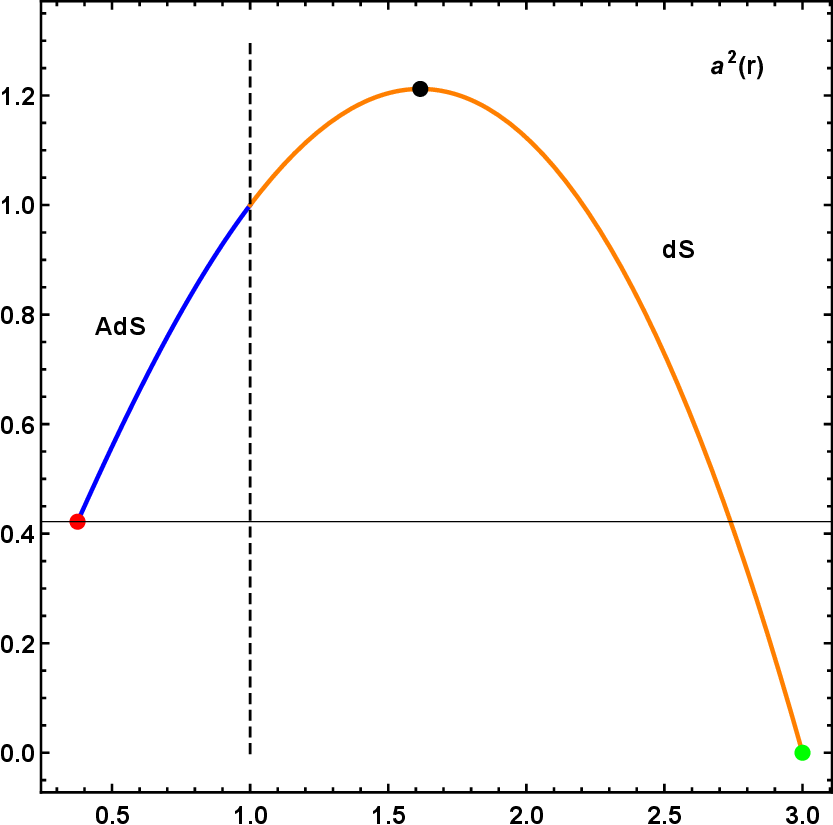,width=0.44\linewidth,clip=} &&&
\epsfig{file=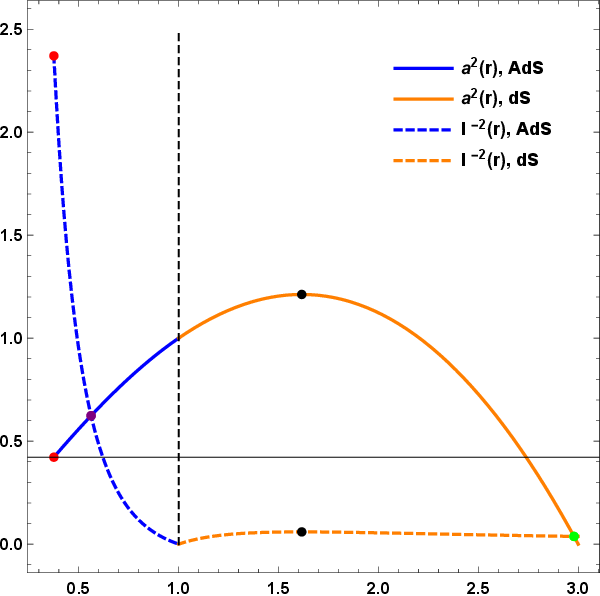,width=0.44\linewidth,clip=}
\end{tabular}
\caption{Comparison of the extremal structure in the AdS and dS cases for vanishing electric charge. The left panel shows the squared rotation parameter $a^2(r)$, where the AdS branch (blue) is confined to $r<M$ and terminates at a geometric endpoint (red point), while the dS branch (orange) extends for $r>M$, develops an ultra-extremal maximum (black point)  and subsequently decreases toward the non-rotating limit at $r=3M$ (green point). The right panel displays both $a^2(r)$ (solid lines) and $l^{-2}(r)$ (dashed lines) for the AdS (blue) and dS (orange) branches. The purple and green points indicate the internal configurations at which the normalized rotational and curvature scales coincide along the AdS and dS branches, respectively, while the red point marks the geometric endpoint of the AdS branch. In all figures, quantities are expressed in units of the mass parameter $M$.}%
\end{figure}%

It is important to note that the  above comparison extends qualitatively to the charged case. The presence of electric charge introduces an additional scale, which modifies the detailed structure of the extremal branches in both AdS and dS geometries, while preserving the key distinction between the two cases.  In particular,  the combined effects of rotation, curvature and charge leads to a richer deformation of the extremal configurations,  without altering the fundamental contrast between boundary-dominated (AdS) and interpolation-type (dS) behavior.

\section{Mass Constraint and Root Structure}

As emphasized in Sec.~III, explicit expressions for the roots of the quartic equation in (\ref{quartic1}) can be obtained using Cardano's method, which relies on a real solution of the associated resolvent cubic equation
\begin{equation}
u^3 - a_2 u^2 - 4 a_0 u - (a_1^2 - 4 a_0 a^2) = 0,
\label{cubic1}
\end{equation}
with coefficients given in (\ref{constcoef}). It is noteworthy that the solution space of this equation admits the special case of a vanishing root, $u = 0$, which in turn leads to a constraint on the mass parameter. Indeed, the roots of the cubic equation satisfy the Vieta relations
\begin{eqnarray}
u_1 + u_2 + u_3 &=& a^2 + l^2 \,, \nonumber \\[2mm]
u_1 u_2 + u_1 u_3 + u_2 u_3 &=& - 4 l^2 (a^2 + q^2) \,, \nonumber \\[2mm]
u_1 u_2 u_3 &=& 4 l^4 \left[ M^2 - (a^2 + q^2)\left(1 + \frac{a^2}{l^2}\right) \right].
\label{rvieta}
\end{eqnarray}
For $u = 0$, the last relation in~(\ref{rvieta}) yields the mass constraint
\begin{equation}
M^2 = (a^2 + q^2)\left(1 + \frac{a^2}{l^2}\right).
\label{constraint}
\end{equation}
This constraint can be interpreted as a saturation condition encoding a balance between rotation, charge  and curvature. It is straightforward to show that, under this condition, the quartic polynomial in (\ref{quartic1}) reduces to
\begin{equation}
 r^4 + l^2 \left( \sqrt{1+\frac{a^2}{l^2}}\, r - \sqrt{a^2+q^2} \right)^2 = 0\,.
 \label{quarticfactor}
\end{equation}
From this form, it follows immediately that the polynomial is strictly positive on the real axis, $f(r) > 0$, and therefore admits no real roots. Consequently, no event or Cauchy horizons can exist in this branch. This conclusion is consistent with the general expressions for the roots in (\ref{horizons}), where the substitution $u = 0$ leads to an unphysical root structure.  Thus, the constraint (\ref{constraint}) defines a branch of solutions without physical horizons. In contrast, under the replacement $l^2 \to -\,l^2$, which maps the above construction to the de Sitter case, the quartic equation (\ref{quartic1}) factorizes into two quadratic polynomials \cite{an1}. As a consequence, the solution space separates into two disjoint branches, each associated with one of the quadratic factors. The roots belonging to different branches do not mix  and within the physical branch there exists a unique extremal configuration. This configuration corresponds to a merger of the outer and cosmological horizons, while excluding the inner--outer horizon merger.

In summary, the algebraic mass constraint leads to qualitatively different root structures in the AdS and dS cases. In AdS case, the real-root structure is eliminated altogether, yielding a branch without physical horizons, whereas in the dS case the quartic equation splits into two disconnected branches admitting a unique extremal merger between the black hole and cosmological horizons.

\section{Conclusion}

A remarkable correspondence exists between a particular class of stationary and axisymmetric spacetimes in general relativity and their constant-curvature counterparts in $f(R)$ gravity, allowing the Kerr--Newman--(A)dS metrics of general relativity to be mapped into the corresponding solutions of $f(R)$ gravity. This correspondence provides a unified framework for investigating the physical properties of both (A)dS and $f(R)$ black holes.  In this paper, we have examined the horizon structure and extremal properties of these black holes, focusing primarily on the anti--de Sitter case within this unified framework.

We analyzed the extremal configurations by solving the underlying system of equations with respect to the squared rotation parameter $a^2$ and the inverse curvature scale $l^{-2}$. Our analysis provides a fully analytic description of the extremal branches and their global structure in the $(a^2,\, l^{-2})$ parameter space, revealing how curvature, rotation, and charge collectively shape the extremal structure. We further showed that the rotating extremal branch remains physically admissible only up to the critical  value of the charge $q_{\max}=M/2$, beyond which the geometry becomes singular and the branch ceases to exist.  Separately examining the uncharged and charged cases, we found that  the extremal branch interpolates between the Kerr--Newman limit and a nonvanishing lower bound of the rotation parameter, thereby implying the existence of a minimal rotation.

Our analysis has also revealed that the extremal branch contains both a geometric endpoint, associated with the breakdown of the spacetime structure and a nontrivial internal scale-matching point  where rotational and curvature effects become comparable. Unlike the geometric boundary, this internal point does not arise from the extremality conditions themselves, but instead reflects an intrinsic balance within the extremal structure. We further showed that electric charge acts as an additional competing scale, enlarging the admissible extremal branch while lowering the corresponding rotation parameter.

We contrasted these results with those obtained for Kerr--Newman--de Sitter black holes in our earlier work \cite{an1}, showing that the extremal structures in AdS and dS spacetimes differ in a fundamental way. While the AdS branch is restricted by a geometric boundary, the dS case exhibits a broader interpolating behavior between maximal and vanishing rotation, reflecting a substantially richer internal structure of the extremal configurations. We also showed that the algebraic mass constraint associated with the quartic horizon equation completely eliminates the real-root structure, leading to a branch of solutions without physical horizons. This behavior differs sharply from the corresponding dS case, where the same algebraic condition gives rise to two disconnected branches and permits a single extremal merger between the black hole and cosmological horizons.

\vspace{-6mm}


\end{document}